\documentclass[preprint,notitlepage,showpacs,preprintnumbers,amsmath,amssymb,superscriptaddress]{revtex4-2}
\usepackage{graphicx,graphics}   % Permite a inser??o de figuras EPS, PDF, JPG ou PNG

\usepackage{graphicx,amsfonts}% Include figure files
\usepackage{epsfig}

\usepackage{dcolumn}% Align table columns on decimal point
\usepackage{bm}% bold math
\usepackage{pslatex}
\usepackage{epsfig}
\usepackage{color}
\usepackage{cancel}
\usepackage{slashed}
\usepackage{amssymb}
\usepackage{amsmath}
\usepackage{hyperref}
\usepackage{orcidlink}
\usepackage[utf8x]{inputenc}
\usepackage{color}
\usepackage{multirow}
\usepackage{slashed}
\usepackage{ulem}
\usepackage{pstricks}
\usepackage[caption=false]{subfig}
\newcommand{\orcid}[1]{\href{https://orcid.org/#1}{\textcolor[HTML]{A6CE39}{\aiOrcid}}}

\begin{document}
%\preprint{IFT-P.xx/2009}
%\preprint{ArXiv:yymm.nnnn}
\title{
What if dark matter is composite of least interacting particles?}
%\maketitleby
\author{Elaine C. F. S. Fortes \orcidlink{0009-0004-5940-3151}}%
\email{elainefortes@unipampa.edu.br}
\affiliation{Universidade Federal do Pampa \\
	Rua Luiz Joaquim de Sá Brito - Promorar – 97.650-000 – Itaqui – RS – Brazil
} 

\author{V. Pleitez  \orcidlink{0000-0001-5279-8438}}%
\email{v.pleitez@unesp.br}
\affiliation{
Instituto  de F\'\i sica Te\'orica--Universidade Estadual Paulista (UNESP)\\
R. Dr. Bento Teobaldo Ferraz 271, Barra Funda\\ S\~ao Paulo - SP, 01140-070,
Brazil
}

%%%%%%%%%%%%%%%%%%%%%%%%%
\date{October 21 2026%16 %7
	%09/07/2026
	}

\begin{abstract}
	
In this work, we investigate the possibility that Dark Matter contains particles that interact with known matter only through gravity, the least interacting particles (LIPs). Specifically, we consider the case of a dark Proca vector field, which does not mix in the kinetic term with any of the vector bosons of the Standard Model if a discrete symmetry ($Z_2$) is imposed. Breaking this symmetry introduces kinetic mixing. Here, we study two realization mechanisms, namely kinetic mixing with the electromagnetic field after electroweak symmetry breaking (Scenario A) and kinetic mixing with the weak hypercharge gauge field (Scenario B). We also consider generalized kinetic terms that violate parity and time-reversal symmetries.

\end{abstract}

%\begin{keyword} 
%KEYWORDS: LIPs, dark photon, parity violation, 
%\file{elsart}, document class, instructions for use
%\PACS 11.30.Hv; %Flavor symmetries
%12.60.Fr; %Extensions of electroweak Higgs sector
%12.15.Ff; %Quark and lepton masses and mixing
%14.60.Pq
%Neutrino mass and mixing,
%\end{keyword}

%\end{frontmatter}

%%%%%%%%%%%%%%%%%555
%\pacs{12.60.Fr; %Extensions of electroweak Higgs sector
%	12.15.-y }

\maketitle
%\newpage

\section{Introduction}
\label{sec:intro}

One of the most important questions in physics and astrophysics is the nature of dark matter (DM) in the universe. At present, its existence has only been confirmed through its gravitational interactions with known matter; for instance, from the rotation curves of galaxies or from cosmological observations~\cite{Khalil:2001fw,Bertone:2004pz,Cirelli:2024ssz,Balazs:2024uyj}. Together with the dark energy and the ordinary matter, DM constitutes an essential component of the standard cosmological model, the usual cosmological model with cold dark matter and dark energy, or cosmological constant, $\Lambda$, i..e, $\Lambda$CDM~\cite{Planck:2018nkj}. Modifications of the theory of general relativity are still candidates, but only on galactic scales~\cite{Skordis:2020eui}. However, despite strong gravitational evidence, the microscopic nature of DM remains elusive.

If the answer to the nature of DM lies in the realm of elementary particle physics, which is a common point of view, then that would already be a clear evidence of new physics beyond the Standard Model (SM). In fact, in the context of a simple extension of this model, possible DM candidates in the universe are sterile neutrinos, although they are very constrained by  invisible $Z$ decay and by direct searches of DM scattering, particularly when they mix with the known active neutrinos~\cite{Cline:2026tkp}. 

From the perspective of elementary particles, the known fields are classified by the interactions they feel. For instance, hadrons experience all four fundamental forces, while charged leptons experience three, and neutrinos only two. A natural question that arises is: why should there not be particles that interact exclusively through gravity? Such particles were called \textit{least interacting particles} (LIPs) in Ref.~\cite{Matsas:1998zm} (see Table~\ref{table}). Here, we do not address the specific mechanism responsible for generating LIPs in the early Universe, whether at the Planck scale or during post-inflationary epochs~\cite{Bastero-Gil:2018uel, Bastero-Gil:2021wsf}, where gravitational interactions are expected to become important. We simply assume their existence.
We are interested only in the possibility of them being dark matter candidates. 

\begin{table}[htbp]
	\centering
	\begin{tabular}{|c|c|c|c|c|}
		\hline\hline
		& \multicolumn{4}{c|}{Interactions} \\ \cline{2-5}
		Particles & Gravity & Weak & EM & Strong \\ \hline\hline
		Quarks          & x & x & x & x \\ \hline
		Charged leptons & x & x & x &   \\ \hline
		Neutrinos       & x & x &   &   \\ \hline
		LIPs            & x &   &   &   \\ \hline
	\end{tabular}
	\caption{Type of particles and their interactions. Here, EM denotes the electromagnetic interaction.}
	\label{table}
\end{table}

LIPs can be scalar, fermionic, or vector fields. We are not interested here in presenting a renormalizable model of LIPs—which is why we regard this as a simplified model—but rather in providing just one example of what could be a very rich dark physics. In any event, pure LIPs can only be detected by their gravitational effects. This is because LIPs couple to all types of matter, but they do so with very weak couplings proportional, in the linear aproximation, to $\sqrt{8\pi G_N}$, where $G_N$ is the Newton's constant. For instance, if, in the future we observed a black hole to evaporate faster than what is predicted by using the standard model of particles, it could indicate that part of the emitted energy is in LIPs form; or even it could be observed through the lensing effect in addition to those mentioned above. 

Then, since we already know about their coupling to matter, we can adjust the LIPs’s mass to obtain the average density of dark matter observed today, which is. approximately $3\times 10^{-30}\,\textrm{g/cm}^3$. Since the volume of the universe is $\approx 1086 \textrm{cm}^3$, then the mass of the total DM in LIPs’s form is approximately 1080 GeV. It means that if the mass of the LIPs is 1 GeV, there must be, at present, this number of LIPs in the Universe. Note that the motivation for hypothesizing the existence of LIPs is independent of any model and, as a candidate for dark matter, it is completely different from any other type of dark matter considered to date. Below we will consider the case of a LIP which is a massive vector field.

For instance, if in the future a black hole is observed to evaporate faster than what is predicted by the SM of particles, it could signal that part of the energy is emitted into LIPs; Similarly, their presence could be observed through the lensing effect in addition to what was mentioned above.
 
Given the gravitational coupling of LIPs, we can perform a simple order-of-magnitude estimate for the total number of particles needed to account for the observed DM density. Taking $\rho_{\rm DM} \sim 3\times10^{-30}\ {\rm g\ cm^{-3}}$ over a characteristic observable volume $V_U \sim 10^{86}\ {\rm cm^3}$, the total DM mass is roughly $M_{\rm DM} \sim 10^{80}\ {\rm GeV}$. For a representative mass $m_{\rm LIP} \sim 1\ {\rm GeV}$, this yields a total population of $N_{\rm LIP} \sim 10^{80}$ within the visible Universe. It means that if the mass of the LIPs is 1 GeV, there must be, at present, this number of LIPs in the Universe.

\section{Generalized kinetic mixing}
\label{sec:pvkm}

Let us introduce a massive dark vector field, $V^\prime_\mu$. This field may mix with one of the neutral vector bosons in the SM, namely, the photon, $A_\mu$, which defines scenario A, or the mixing of the vector LIp occurs with the weak hypercharge gauge boson $B$, which we call scenario B. Scenario B differs from scenario A in that left-handed neutrinos are included.

If a discrete $Z_2$ symmetry is imposed—under which LIPs are odd and all SM particles are even—the dark vector becomes strictly stable, interacting with itself and ordinary matter solely through gravity. Under these conditions, the numerical estimates presented above hold, and DM would remain observable exclusively via its gravitational footprint."

However, if we do not impose the $Z_2$ symmetry, the dark vector field, $V_\mu$, kinetically mixes with one of the gauge fields in the electroweak SM. This mixing can occur with the $U(1)_{em}$, or with the vector boson related to $U(1)_Y$, depending on whether it takes place after or before spontaneous symmetry breaking. First examples of this type of vector field were introduced in Refs.~\cite{Galison:1983pa}, and such fields have been studied as a DM candidate. See Refs.~\cite{Cline:2024qzv,Bauer:2018onh,Roy:2026sek,
Griffith:2026hdi,he:2026, Fortes:2015qka, Fortes:2017kca} and references therein.

Here we detail the case where the dark vector $V^\prime_\mu$ kinetically mixes with the photon (Scenario A). Scenario B is obtained from Scenario A by replacing the field $A$ with the field $B$ and the corresponding interactions including neutrinos. 

The total Lagrangian, which includes parity-violating terms, is given by 
\begin{equation}
\mathcal{L}=\mathcal{L}_{SM}+\mathcal{L^\prime}_{kin}
\label{lgeral}
\end{equation}
where the second term involves the interaction between the vector fields $A^\prime$ (or $B^\prime$) and the dark vector $V^\prime_\mu$.
\begin{eqnarray}
\mathcal{L}^{\prime}_{kin}&=&-\frac{1}{4}F^\prime_{\mu\nu}F^{\prime\,\mu\nu}+\frac{M_{V}^{2}}{2}V^{\prime}_{\mu}
V^{\prime \mu}
\!-\!\frac{1}{4}V^{\prime}_{\mu\,\nu}V^{\prime\mu\nu}+
\frac{g_{VA}}{2}V^{\prime}_{\mu\,\nu}F^{\prime\mu\nu}
\nonumber\\& -&\frac{\kappa}{4} F^\prime_{\mu\nu}\tilde{F^\prime}^{\mu\nu} -g_{PV}V^\prime_{\mu\nu}\tilde{V}^{\prime\,\mu\,\nu}-\frac{g^\prime_{PV}}{2}V^\prime_{\mu\nu}\tilde{F}^\prime_{\mu\nu}
-\frac{g^{\prime\prime}_{\,PV}}{2}\tilde{V}^\prime_{\mu\nu}F^{\prime\,\mu\nu},
\label{k1}
\end{eqnarray}
where  $\tilde{X}^\prime_{\mu\nu}=(1/2)\epsilon_{\mu\nu\alpha\beta}X^{\prime\alpha\beta}$, where $X^\prime_{\mu\nu}=F^\prime_{\mu\nu}$, or $X^\prime_{\mu\nu}=V^\prime_{\mu\nu}$. The primes denote the fields before the parity-conserving term is diagonalized by the transformation below in Eq.~(\ref{k2}). The kinetic mixing may, or may not, be generated by gravitational interactions; hence, the parameters $g_{VA}, g_{PV}, g^\prime_{PV}$ and $g^{\prime\prime}_{PV}$ may be $\ll O(1)$ or $O(1)$. The dark photon mass will be assumed to be non-zero. Here, we will not concern ourselves with how this mass is generated.

Making the transformation
\begin{equation}
\left(\begin{array}{c}
V^\prime \\A^\prime\end{array}\right)=\left(
\begin{array}{cc}
\frac{1}{\sqrt{1-g^2_{VA}}} & 0 \\
\epsilon_{VA} & 1\\
\end{array}\right)
\left(\begin{array}{c}
V\\ A
\end{array}
\right),
\label{k2}
\end{equation}
where 	$\epsilon_{VA}=\frac{g_{VA}}{\sqrt{1-g^2_{VA}}}$, we eliminate the kinetic mixing conserving parity and modify the parameters in Eq.~(\ref{k1}). In scenario B, we make $F^\prime_{\mu\nu}\to B^\prime_{\mu\nu}\,A^\prime_\mu\to B^\prime_\mu$, and so on, taking into account the weak interaction also with $B$.

When we use the transformation in Eq. (\ref{k2}), the Lagrangian $\mathcal{L}^\prime_{kin}$ becomes diagonal in the parity conserving sector
\begin{eqnarray}
\mathcal{L}_{kin}&=& \frac{1}{2}\bar{M}_{V}^{2}V_{\mu}V^{ \mu}  -\frac{1}{4}V_{\mu\nu}V^{\mu\nu}-\frac{1}{4}F_{\mu\nu}F^{\mu\nu}\nonumber \\&-&\frac{\kappa}{4}F_{\mu\nu}\tilde{F}^{\mu\nu}-\varepsilon_{PV}
V^{\mu\nu}\tilde{V}_{\mu\nu}-\kappa_{PV} \tilde{V}^{\mu\nu}F_{\mu\nu}
\label{k3}
\end{eqnarray}
with $\bar{M}^2_V=M^2_V/(1-g^2_{VA})$, and we have taken into account that $V^{\mu\nu}\tilde{F}_{\mu\nu}=\tilde{V}^{\mu\nu}F_{\mu\nu}$, and we have defined
\begin{eqnarray}
&&\varepsilon_{PV}=\frac{1}{1-g^2_{VA}}\left(\frac{\kappa}{4}\,g^2_{VA}+g_{PV}
+\frac{g_{VA}}{2}(g^\prime_{PV}+g^{\prime\prime}_{PV})\right),
\nonumber \\&&
\kappa_{PV}=\frac{1}{2\sqrt{1-g^2_{VA}}}\left(\kappa\, g_{VA}+g^\prime_{PV}+g^{\prime\prime}_{PV}\right)
\label{def}
\end{eqnarray} 

Notice that, after the field redefinition in Eq. (3), the parity-conserving kinetic mixing term $V_{\mu\nu}F^{\mu\nu}$ is eliminated. The remaining generalized terms involve the dual field-strength tensors, namely $F_{\mu\nu}\widetilde{F}^{\mu\nu}$, $V_{\mu\nu}\widetilde{V}^{\mu\nu}$, and $\widetilde{V}_{\mu\nu}F^{\mu\nu}$. In terms of the electric and magnetic fields, the mixed dual contraction contains the combination $\vec{E}_D\cdot\vec{B}+\vec{E}\cdot\vec{B}_D$, up to convention-dependent signs and factors. These terms characterize the parity- and time-reversal-violating sector of the theory. The physical properties of the dark vector under $P$ and $T$ are not necessarily the same as those of the photon.

We can introduce scalars or fermionic LIPs. For instance, a real scalar $\phi$ and two Dirac fermions, $\eta$. These fields are purely LIPs if, for instance, we introduce the discrete symmetries $Z_2\otimes Z_3$ under which, $V_\mu,\phi,\eta_1,\eta_2$ transform under $Z_2$ as $-1,+1,-1,+1$ and under $Z_3$ as $\omega,\omega^{-1},\omega,\omega$, respectively. Additionally, all the SM particles are even under both discrete symmetries. Even fermionic LIPs have to satisfy the Tremaine-Gunn contraints~\cite{Tremaine:1979we}, which yield a lower bound of $M_\eta>100-400$ eV. These limits arise because Pauli's exclusion principle limits the number of fermions that can fit into a given volume. The strongest constraints come from  dwarf spheroidal galaxies. On the other hand, for bosonic ones, the constraints stem from the requirement that their wavelength must be smaller than a typical size of dwarf galaxies, i.e., on the order of a thousand light-years, which translates to a lower limit mass of approximately  $10^{-22}$ eV, for a distance of 1000 light-years. 

Note that, unlike other particles that experience only gravitational interaction, our LIPs arise in a model-independent manner, as illustrated in Table~\ref{table}. They simply represent a different type of particle. Moreover, it should also be noted that the mixing of Abelian vector bosons in the kinetic term requires no ad hoc justification, since all the symmetries of the theory permit it, unless we introduce appropriate discrete symmetries.

%%%%%%%%%%%%%%%%%%%%%%%%%%%%%%%%%%%%%%%%%%%%%%%%%%%%%%%%%%%%%%%%%%%%%%%%

\section{Consequences of the kinetic mixing}
\label{sec:consequences}

%%%%%%%%%%%%%%%%%%%%%%%%%%%%%%%%%%%%%%%%%%%%%%%%%%%%%%%%%%%%%%%%%%%%%%%%%%%%%
Particles that experience only gravitational interactions may have been produced and possibly brought into equilibrium during the very early Universe, when the gravitational effects were much stronger than at later cosmological epochs. As the universe expanded and the gravitational interaction became comparatively weak, LIPs were expected to decouple very early from the thermal bath. If only gravitational interactions are allowed for LIPs, their subsequent cosmological evolution is governed essentially by the expansion of the universe. So, provided  that their primordial abundance is appropriately established, the present-day density of LIPs may account for a part or all of the observed DM density, as we discussed above.

However, in the case of a vector field, the kinetic mixing with, e.g., the photon may induce a reheating period. From that point onward, one needs to evaluate whether the resulting relic density can explain the current DM content. The most intriguing scenario, however, would be if the dark vectors were just undergoing decoupling at the present epoch.

The kinetic mixing terms in Eq~(\ref{k1}) are not necessarily induced by gravitational interactions. Since such mixing is fundamentally allowed in quantum field theory, the parameters in Eq.~(\ref{k3}) may be larger than $\sqrt{8\pi G_N}$. In particular the factor $1/\sqrt{1-g^2_{VA}}$ may induce an enhancement of the respective interactions. 

\noindent \textbf{Scenario A} 

As a consequence of the transformation in Eq.~(\ref{k2}), charged particles may interact with $V_\mu$: 

\begin{equation}
	\mathcal{L}_{\bar{f}fV}\to -e\,\epsilon_{VA}\left(\sum_f\,Q_f\,\bar{f}\,\gamma^\mu f\right)\,V_\mu
	\label{ncs2}
\end{equation}  
$e$ is the electric charge of the positron, and $f$ run over all known charged fermions. This is scenario A, in which dark photon interactions with Standard Model particles occur solely through electric charge; neutrinos are automatically excluded. 
 
The other neutral interactions are 

\begin{eqnarray}
\mathcal{L}_{NC}&=& -e\sum_f Q_f\bar{f}\gamma^\mu f\, A_\mu-\frac{g}{2c_W}\sum_f\bar{f}\gamma^\mu (g^f_V-\gamma_5g^f_A)f\,Z_\mu,
%\nonumber \\&-& \frac{g^\prime}{2}\epsilon_{VB}\,\sum_f\left[Y^f_L %\bar{f_L}\gamma_\mu f_L+Y^f_R \bar{f_R}\gamma_\mu f_R\right]\!\!V_\mu,\;\;
\label{nc1}
\end{eqnarray}
where $g$ and $g'$ are the usual $SU(2)_L$ and $U(1)_Y$ gauge coupling constants, respectively,  $f=e,\nu,u,d,...b,t$, and
\begin{eqnarray}
&&g^f_V=T_3^f-2Q^fs^2_W,\; g^f_A=T^f_3,\quad (a) \nonumber \\&&  Y^{l,\nu}_L=-1,\; Y^{u,d}_L=1/3,\; Y^l_R=-2,\; Y^u_R=4/3,\; Y^d_R=-2/3 \quad (b).
%g^f_V+g^f_A,\quad Y^f_R=g^f_V-g^f_A,\quad (b).
\label{fgs} 
\end{eqnarray}

The  neutrinos are  included in Scenario B.

Thus, in Scenario A, $V_\mu$ may decay into all kinematically allowed charged fermions, with partial decay widths given by
\begin{equation}
\Gamma(V\to f\bar{f})=\frac{N^f_ce^2Q^2_f\epsilon^2_{VA}M_V}{12\pi}\,\left(1+\frac{2m^2_f}{M^2_V}\right)\,\sqrt{1-\frac{4m^2_f}{M^2_V}}.
\label{width}
\end{equation}

\noindent \textbf{Scenario B}

In this case, kinetic mixing occurs with the vector boson associated with the weak hypercharge $U(1)_Y$ factor, denoted, as usual, as $B_\mu$, and the neutral interactions are different.

Now we have
\begin{equation}
\mathcal{L}_B=
-\left[
g^f_L\,\bar f_L\gamma^\mu f_L+
g^f_R\bar f_R\gamma^\mu f_R
\right]V_\mu,
\end{equation}
where
\begin{eqnarray}
&&g_L^f=\frac{g^\prime\epsilon_{VB}}{2}Y_L^f,\qquad g_R^f=\frac{g^\prime\epsilon_{VB}}{2}Y_R^f,
\qquad
r_f=\frac{m_f^2}{M_V^2}\nonumber \\&&  Y^{l,\nu}_L=-1,\; Y^{u,d}_L=1/3,\; Y^l_R=-2,\; Y^u_R=4/3,\; Y^d_R=-2/3.
\end{eqnarray}

The partial decay width into a pair of fermions is

\begin{equation}
\Gamma_B(V\to f\bar f)=
\frac{N_c^f g^{\prime\,2}\epsilon_{VB}^2M_V}{96\pi}
\sqrt{1-4r_f}
\left[\left((Y_L^f)^2+(Y_R^f)^2\right)(1-r_f)+6Y_L^fY_R^f\,r_f\right].
\end{equation}

\begin{equation}
\Gamma_B(V\to f\bar f)\simeq\frac{N_c^f g^{\prime\,2}\epsilon_{VB}^2M_V}{96\pi}
\left[(Y_L^f)^2+(Y_R^f)^2\right].
\end{equation}

For active neutrinos $Y^\nu_L=-1,\,Y^\nu_R=0$ we have

\begin{equation}
\Gamma_B(V\to\nu\bar\nu)=\frac{g^{\prime\, 2}\epsilon_{VB}^2}{96\pi}\,M_V.
\end{equation}

%\noindent \textbf{daqui pra frente volta ao caso A?}\\

\section{Analysis}
\label{sec:analisis}

After the first decoupling, i.e., before kinetic mixing becomes relevant, the dark vector is effectively decoupled from the Standard Model plasma. Once kinetic mixing is established, interactions with SM particles may bring the dark vector back into thermal contact, potentially leading to a second decoupling at a later cosmological epoch.

In Scenario A, we have the total width
\begin{equation}
\Gamma^{tot}_{V;A}=\sum_{f\,charged}\Theta(M_V-2m_f)\Gamma_A(V\to f\bar{f})
\end{equation}
where only charged fermions contribute, and in Scenario B:
\begin{equation}
	\Gamma^{tot}_{V;B}=\sum_{all f}\Theta(M_V-2m_f)\Gamma_B(V\to f\bar{f})
\end{equation}
where $f$ includes neutrinos.

Assuming $H(t)=H_0$, with the present Hubble constant given by $H_0\sim 2.7\times10^{-18}\,\textrm{s}^{-1}\approx 1.5\times10^{-42}$ GeV, the dark vector is cosmologically long-lived if $\Gamma_V^{\mathrm{tot}}<H_0$, or, equivalently, if its lifetime satisfies $\tau_V=(\Gamma_V^{\mathrm{tot}})^{-1}>H_0^{-1}$.

%\begin{equation}
%\Gamma(V\to f\bar{f})\vert_{M_V<2m_b}\approx \frac{N_c\alpha %Q^2_b}{3}\epsilon^2_{VA}2m_b < 2.7\times10^{-42}\,\textrm{GeV}.
%\label{desaV}
%\end{equation}
%or, for $N_c=3$ and $Q^2_b=1/9$, we have
%\begin{equation}
%\epsilon^2_{VA}>\frac{9H_0}{8\alpha m_b}.
%\label{desV2}
%\end{equation}

The total decay width determines whether the dark vector can survive over cosmological timescales. Since the lifetime of the vector is given by $\tau_V=(\Gamma_V^{\rm tot})^{-1}$, the condition $\Gamma_V^{\rm tot}<H_0$ corresponds approximately to $\tau_V>H_0^{-1}$ and therefore defines the
cosmologically long-lived region. Conversely, if
$\Gamma_V^{\rm tot}>H_0$, the vector lifetime is shorter than the Hubble timescale and the particle decays before the present epoch.

For a given mass $M_V$, this requirement establishes an upper limit on the kinetic mixing parameter, enforcing a sufficiently small $\epsilon_{VA}$ to ensure cosmological stability.

\begin{figure}[htbp]
\centering
\includegraphics[width=\textwidth]{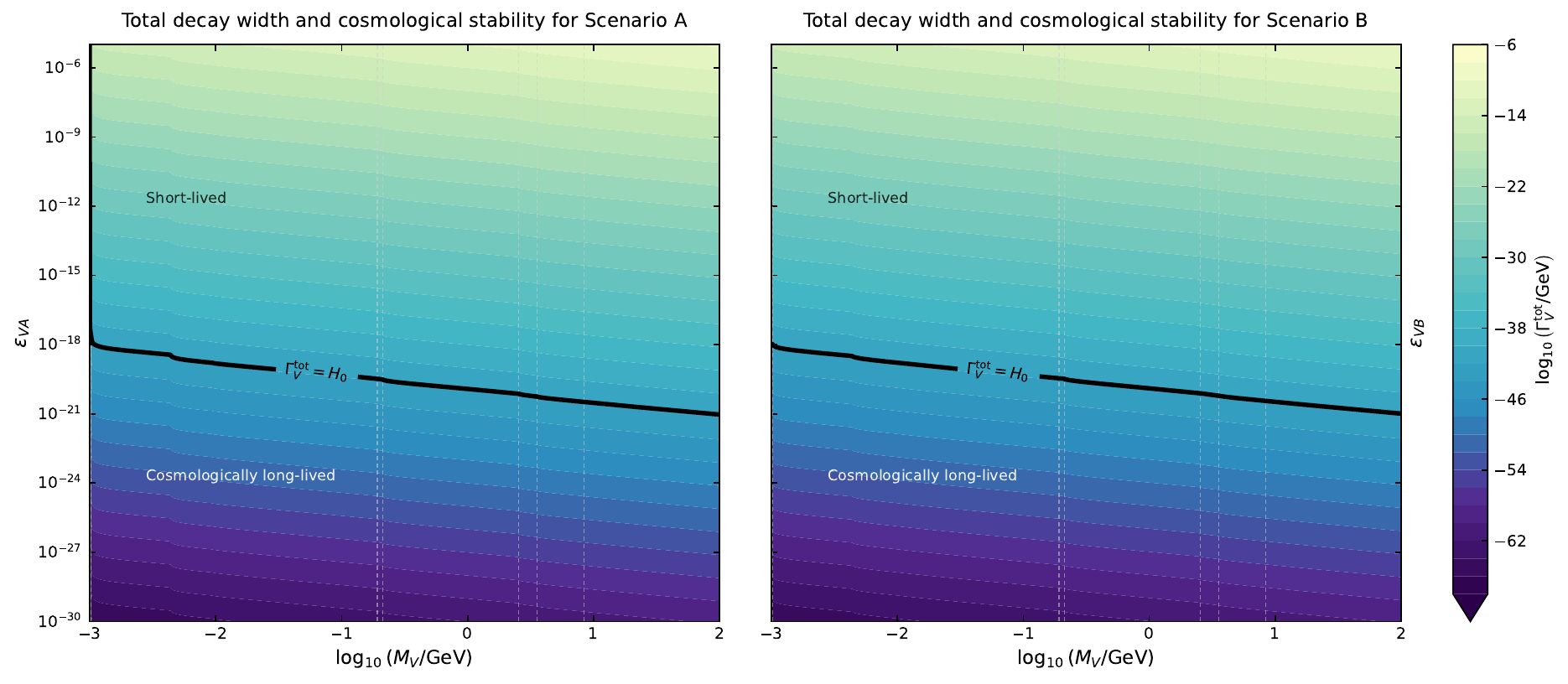}
\caption{Contour plots of the total decay width of the dark vector boson for the two kinetic-mixing scenarios investigated in this work. The left panel corresponds to Scenario A, with kinetic mixing with the photon, and is shown in the $(\log_{10}(M_V/\mathrm{GeV}),\,\log_{10}\epsilon_{VA})$ plane. The right panel corresponds to Scenario B, with kinetic mixing with the hypercharge gauge boson, and is shown in the $(\log_{10}(M_V/\mathrm{GeV}),\,\log_{10}\epsilon_{VB})$ plane. The color scale represents $\log_{10}(\Gamma_V^{\mathrm{tot}}/\mathrm{GeV})$, and the solid black curve indicates the cosmological stability condition $\Gamma_V^{\mathrm{tot}}=H_0$, separating the cosmologically long-lived region $(\Gamma_V^{\mathrm{tot}}<H_0)$ from the unstable region $(\Gamma_V^{\mathrm{tot}}>H_0)$. The dashed vertical lines mark the selected kinematic thresholds $M_V=2m_f$ for the massive fermionic final states, where new decay channels become available. In both panels, the displayed thresholds extend from the electron to the $b$-quark. In Scenario B, the three active neutrino channels are also included; since neutrino masses are neglected, they do not generate a finite kinematic threshold in the mass range shown. Their contribution increases the total decay width and shifts the cosmological-stability boundary toward smaller values of $\epsilon_{VB}$.}
\label{fig:1A}
\end{figure}

In Figure \ref{fig:1A}, the color scale represents ($\log_{10}(\Gamma_V^{\rm tot}/{\rm GeV}$)), and the solid curve indicates the condition $\Gamma_V^{\rm tot}=H_0$. The region below this curve corresponds to $\Gamma_V^{\rm tot}<H_0$, where the lifetime of the dark vector is longer than the Hubble timescale. Therefore, the particle can therefore be regarded as cosmologically long-lived. Conversely, in the region above the curve, ($\Gamma_V^{\rm tot}>H_0$), the decay timescale is shorter than the Hubble timescale, so that the vector cannot remain stable over cosmological times.

The dashed vertical lines indicate the kinematic thresholds ($M_V=2m_f$), at which new decay channels ($V\to f\bar f$) become accessible and begin to contribute to the total width. In the mass range considered here, these include the channels involving (e,$\mu$,$\tau$) and ($u,d,s,c,b$). Decays into ($W^+W^-$) and ($t\bar t$) are not included because the mass range considered here lies below the corresponding kinematic thresholds.

\section{Possible Detection}
\label{sec:detection}

By introducing a real scalar field $\phi$~\cite{Garcia-Garcia:2026nzy}, and not considering the $Z_2\otimes Z_3$ symmetries, the effective interactions Lagrangian takes the form:
\begin{equation}
-\mathcal{L}_{\phi2}=\phi\left(\frac{c_1}{\Lambda_1}V^\prime_{\mu\nu}
\tilde{V^\prime}^{\;\mu\nu}+\frac{c_2}{\Lambda_2}\,F^\prime_{\mu\nu}
\tilde{V^\prime}_{\mu\nu} %	+\frac{c_3}{\Lambda_3}\tilde{F}^\prime_{\mu\nu} V^\prime_{\mu\nu}
\right)
\label{chiV}	
\end{equation}
where $c_i$ are real, or complex parameters and the field $\phi$ may be a pseudoscalar. This real scalar(pseudoscalar) field may constitute an inflaton-like LIP as considered in Ref.~\cite{Bastero-Gil:2021wsf} in which a mechanism for production of the dark photon at the end of inflation was proposed by considering only the $c_1$ interaction in Eq.~(\ref{chiV}).

If the kinetic mixing occurs after spontaneous symmetry breaking, the real boson may be the Higgs field, $H^0$, i.e., $\phi\equiv H^0$. At the LHC, such interactions may be searched for through missing-transverse-energy, for process as $H^{0*}\to X+$photon+missing energy. However, in this case one must establish whether the missing-energy signal is associated with the cosmological DM component in cosmology and astronomy by its gravitational effect. 

Signatures of LIPs could also manifest in strong-gravity or high-density astrophysical environments, ranging from the Sun and neutron stars to black holes. In the latter, the presence of LIPs would modify Hawking radiation relative to SM predictions. However, in this case, the linear approximation is no longer valid.

Anapole moments offer a viable channel for DM detection. Their interactions are short-ranged effective interactions with highly suppressed couplings. Notably, the anapole moment represents the sole electromagnetic multipole permitted for Majorana fermions. In our framework, the presence of parity-violating interactions naturally yields anapole-type couplings~\cite{Zeldovich:1958lnu, zeldovich58b, Haxton:2001zq, Herrera:2024ysj}, 
\begin{equation}
-\mathcal{L}_{a_V}=\frac{4\pi a_V}{c\Lambda^2\sqrt{1-g^2_{VB}}}\,
\varepsilon^{\mu\,\nu\,\alpha\,\beta}V_\nu  (\partial_\alpha  V_\beta)\,J_\mu^{ext}
\label{ana1}
\end{equation}	
where $\Lambda\sim m_V$ and $a_V$ is a dimensionless parameter. We have used Maxwell's equation $\partial^\alpha F_{\alpha\nu}=\frac{4\pi}{c} J_\nu^{ext}$, where $J_\nu^{ext}$ is an external electric current.

Anapole moments are naturally suppressed and do not appear at tree level; they are loop-induced effective operators that describe contact interactions, contributing only where the external current is non-zero. Because the anapole operator is parity-odd, it cannot be generated in a purely parity-conserving theory such as QED; its presence requires an underlying parity-violating sector. For instance, the $Z$ boson of the SM acquires an anapole moment of the order of $10^{-33}e\cdot$cm. Neutrinos may also have anapole moments and in the SM context, they are of the order of $10^{-34}e\cdot$cm.
The magnetic anapole interaction induces a small atomic recoil which it is expected to be observed in appropriate experiments such as XENONnT~\cite{XENON:2025vwd} and LUX-ZEPLIN~\cite{LZ:2024zvo}. 

\section{Conclusions}
\label{sec:con}

We investigated a class of DM candidates based on LIPs, focusing on massive vector fields and the phenomenological impact of kinetic mixing with the neutral SM gauge sector. The concept of LIPs does not rely on a specific underlying theory. Rather, it is motivated by the observation that known particles are distinguished by the interactions they experience. While some particles participate in all four fundamental interactions, others experience only three or two. It is therefore natural to consider the possibility of particles that interact only gravitationally. Such particles define the class we refer to as least interacting particles.

The existence of LIPs is model-independent and its spectrum may comprise both scalar and fermionic degrees of freedom. In the scalar case, unwanted interactions with the SM Higgs doublet can be prevented by imposing appropriate discrete symmetries. For dark fermions, mixing with visible matter is naturally forbidden provided that they carry different quantum numbers such as electric charge from the SM quarks and leptons.

Kinetic mixing breaks the absolute stability of the vector field by inducing couplings to SM fermions enabling the decays $V \to f\bar{f}$. To assess its viability as a DM candidate, we evaluate its total decay width against the Hubble parameter today, where $\Gamma_V^{\rm tot} = H_0$ defines the boundary between cosmologically long-lived and unstable regions. 
For $\Gamma_V^{\rm tot}<H_0$, the vector can survive over cosmological timescales, whereas for $\Gamma_V^{\rm tot}>H_0$ its lifetime is shorter than the Hubble timescale. For a given vector mass, the stability condition therefore imposes an upper bound on the kinetic-mixing parameter. In Scenario B, kinetic mixing with the hypercharge gauge boson opens additional decay channels, including neutrino final states, increasing the total width and tightening the stability bound.

In summary, we investigated a DM candidate that interacts with the SM exclusively through gravitational couplings or, in the case of a vector field, via kinetic mixing. Consequently, direct detection relies either on gravitational signatures or on loop-induced electromagnetic form factors, such as anapole moments enabled by kinetic mixing. This framework naturally accommodates scalar and fermionic dark sector fields, providing a flexible and model-independent approach to DM phenomenology.

\section*{Acknowledgments}

Elaine C. F. S. Fortes acknowledges the Federal University of Pampa for its institutional support during the development of this research. V. P. would like to thank G. E. A. Matsas for useful discussions.

\end{document}